\documentclass[structabstract]{aa}  
\usepackage{graphicx}
\usepackage{txfonts}
\begin{document}
   \title{High-resolution radio emission from RCW~49/Westerlund~2}

   \author{P. Benaglia\inst{1,2}\fnmsep\thanks{member of Conicet},
           B. Koribalski\inst{3},
           C. S. Peri\inst{1,2},
           J. Mart\'{\i}\inst{4},
           J. R. S\'anchez-Sutil\inst{4},
           S. M. Dougherty\inst{5},
           \and
           A. Noriega-Crespo\inst{6}
          }

   \institute{Instituto Argentino de Radioastronom\'{\i}a, 
CCT-La Plata (CONICET), C.C.5, (1894) Villa Elisa, Argentina\\
              \email{pbenaglia@fcaglp.unlp.edu.ar}
     \and
           Facultad de Ciencias Astron\'omicas y Geof\'isicas, UNLP, 
Paseo del Bosque s/n, (1900) La Plata, Argentina
     \and 
          Australia Telescope National Facility, CSIRO Astronomy \& 
Space Science, PO Box 76, Epping, NSW 1710, Australia
     \and 
          Escuela Polit\'ecnica Superior de Ja\'en, Universidad de Ja\'en, 
Campus Las Lagunillas, Edif. A3, 23071 Ja\'en, Spain\\
  \email{jmarti@ujaen.es, jrssutil@ujaen.es}
     \and
         National Research Council of Canada, Herzberg Institute 
for Astrophysics, Dominion Radio Astrophysical Observatory, 
P.O. Box 248, Penticton, BC V2A 6J9, Canada
     \and
        Infrared Processing and Analysis Center, California Institute of 
Technology, Pasadena, CA 91125, USA  
         }

   \date{Received: May 28, 2013; accepted August 16, 2013}

  \abstract
   {}
   {The HII region RCW~49 and its ionizing cluster form an extensive,
     complex region that has been widely studied at infrared (IR) and
     optical wavelengths. The Molonglo 843
     MHz and Australia Telescope Compact Array data at 1.4 and 2.4 GHz
     showed two shells. Recent high-resolution IR imaging revealed
     a complex dust structure and ongoing star formation. 
     New high-bandwidth and high-resolution data of the
     RCW~49 field have been obtained 
     to survey the radio emission at arcsec scale and investigate the 
     small-scale features and nature of the HII region.}
   {Radio observations were collected with the new 2-GHz bandwidth
     receivers and the CABB correlator of the Australia Telescope
     Compact Array [ATCA], at 5.5 and 9.0 GHz.  In addition, archival
     observations at 1.4 and 2.4 GHz have been re-reduced and
     re-analyzed in conjunction with observations in the optical,
     IR, X-ray, and gamma-ray regimes.}
    {The new 2-GHz bandwidth data result in the most detailed radio
      continuum images of RCW~49 to date.  The radio emission closely
      mimics the near-IR emission observed by Spitzer, showing pillars
      and filaments. The brightest continuum emission comes from the region
      known as the bridge. 
      The overall flattish spectral index is typically consistent with a
      free-free emission mechanism. However, hints of nonthermal
      components are also present in the bridge. 
      An interesting jet-like structure surrounded by a bubble feature
      whose nature is still unclear
      has been discovered close to the Westerlund~2 core.
      Two apparent bow shocks and a number of discrete
      sources have been detected as well in the surroundings of RCW~49.
      In addition, we also report on and discuss the possible detection 
      of a hydrogen recombination line.}
  {The radio results support an association between the cm continuum 
    and molecular emission. The detection of the radio recombination line
    kinematically favors a RCW~49 distance of 6--7 kpc.  If
    the negative spectral indices measured at the bridge should be confirmed
    to be caused by synchrotron emission, we  propose a scenario where 
    high-energy emission could be produced. Finally, the newly discovered
    jet-like structure appears to be an
    intriguing source that deserves a detailed study by itself.}


\authorrunning{Benaglia et al.}
\titlerunning{RCW~49 and Westerlund~2}

   \maketitle
%


\section{Introduction}

RCW~49 has been one of the most often studied HII regions of the southern sky
since its discovery in the sixties (Rodgers et al. 1960) in H$\alpha$
emission. As part of the comprehensive study of more than 200 HII regions, 
Goss \& Shaver (1970) were the first to characterize and image the
source at 5 GHz.
The region was extensively studied in radio recombination
lines with single-dish instruments (see Caswell \& Haynes 1987 and 
references therein). Until now, the low declination of RCW~49, which
makes it inaccesible for northern interferometers,
combined with the 
very large angular extent ($\sim 40' \times 50'$) which made observing 
excessively time-consuming, precluded arcsec
resolution cm-radio observations.

Images
obtained at IR wavelengths (Whitney et al. 2004) in the area of the
Spitzer-GLIMPSE program (Benjamin et al. 2003) pointed to the cluster
Westerlund~2 [Wd2] as the ionizing agent. This cluster is among the
five super star clusters known in the galaxy (Johnson 2005), with
stellar densities above 10$^4$ stars per pc$^{3}$. Piatti
et al. (1998) derived an age of 2--3 Myr and an average visual
absorption of 5 mag. Ascenso et al. (2007) estimated a total stellar
mass up to $\sim 7\times 10^4$ M$_\odot$, assuming a distance of 2.8
kpc.  Over five decades, distance values derived for both the HII
region and the super star cluster presented huge scatter, ranging from
2.5 to 8 kpc. The distance to the region remains an open topic.

A $\sim15''$ by $15''$ field centered on Westerlund~2 was surveyed in
X-rays by Tsujimoto et al. (2007) and Naz\'e et al. (2008) with
Chandra. Tsujimoto and collaborators detected X-ray emission from the early-type
stars, cataloged hundreds of cluster members, mostly
pre-main-sequence and early-type stars, and identified about 30 new OB
star candidates.  This led them to derive a constraint of 2 -- 5 kpc
for the distance to the cluster, based on the mean luminosity of
T-Tauri stars.  Naz\'e et al. (2008) measured and explained the
brightness variability of the eclipsing, colliding-wind binary WR 20a
and found that faint, soft and diffuse emission pervades the field of
view, but no clear structure could be identified. Fujita et al. (2009)
analyzed Suzaku observations of Westerlund~2 and found diffuse
X-ray emission consisting of thermal and maybe one nonthermal
component\footnote{We used the convention $S_{\nu} \propto
\nu^\alpha$, where $S_\nu$ is the flux density at a frequency
$\nu$ and $\alpha$ is the spectral index.}.

\begin{figure*}[!t]
\centering
\includegraphics[width=0.75\textwidth,angle=0]{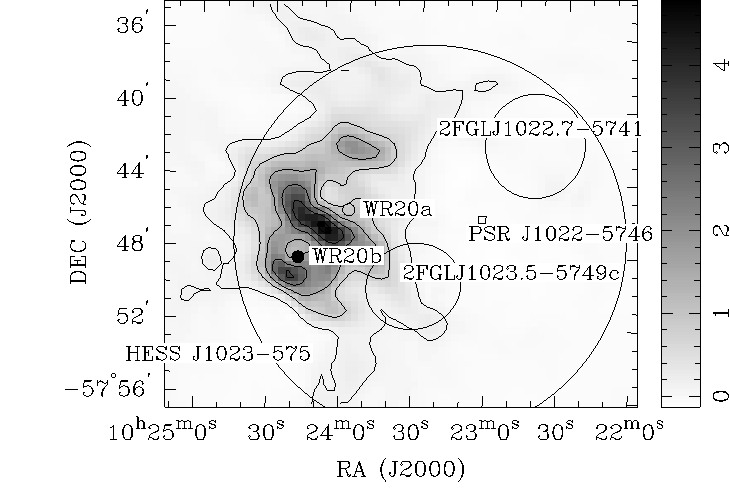}
\caption{Field of RCW~49. Emission at 843 MHz (Whiteoak \& Uchida 1997)
is presented in contours and grayscale (in units of Jy). 
The positions of the two very
bright stars WR 20a (member of Westerlund~2) and WR 20b and of the 
high-energy sources 2FGLJ 1022--5741.7, 2FGLJ 1023.5--5749c, 
HESS J1023--575, and PSR J1022--5746 are shown.}
\label{finder-chart}%
\end{figure*}

\begin{figure*}[!t]
\centering
\includegraphics[width=0.75\textwidth,angle=0]{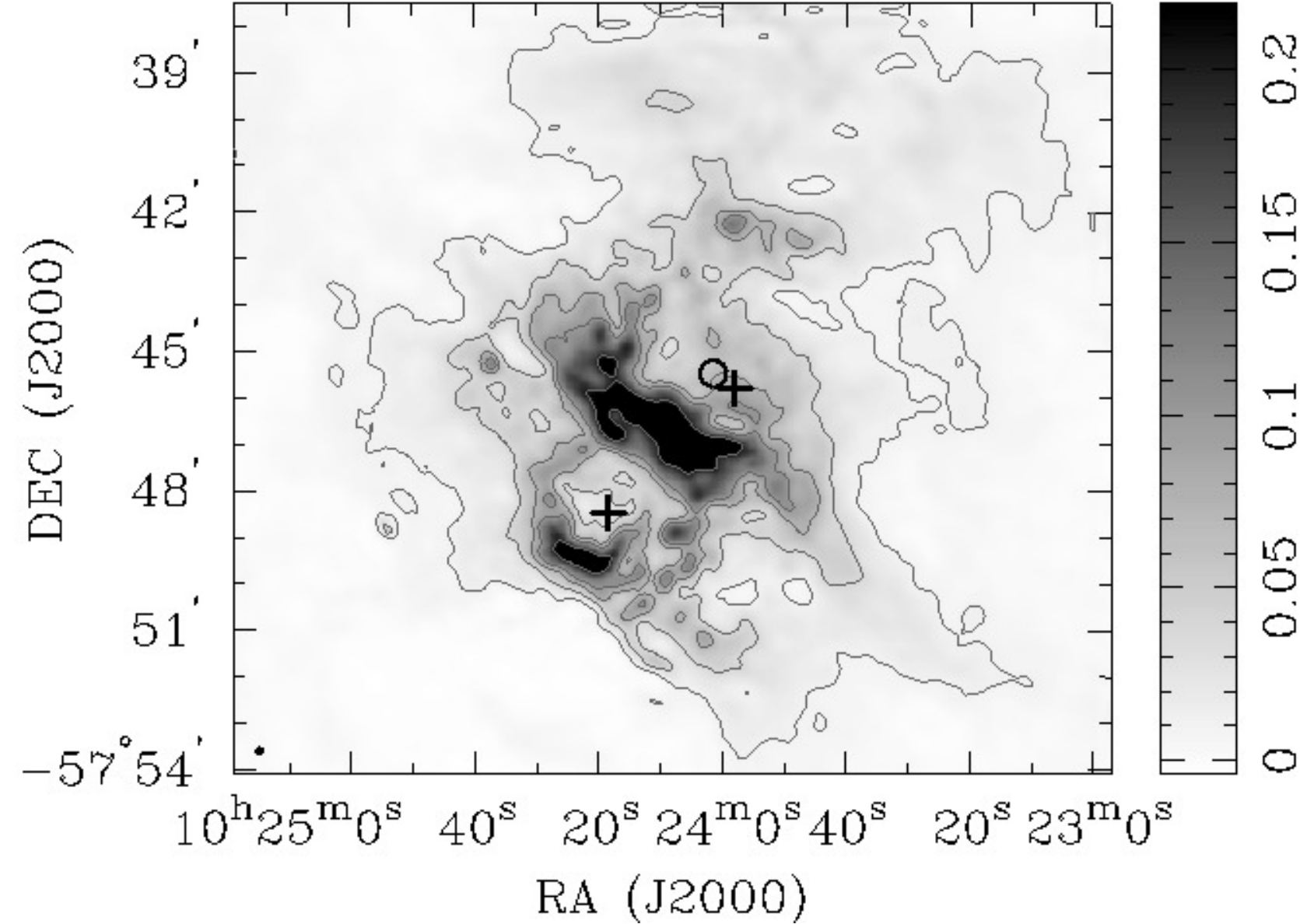}
\caption{Image of the reprocessed 1.4-GHz data taken by WU97 of RCW~49
(see Sec.~\ref{subsec:archivedata}). The position of Westerlund~2 is
marked by the circle. The two very bright stars WR 20a (member of
Westerlund~2) and WR 20b are indicated by the two crosses.  The
bridge is clearly seen as the bright bar of emission in the center
of the image to the southeast of Westerlund~2. Farther to the
southeast is WR20b, surrounded by a circle of emission, the southern
ring as identified by WU97. There is a suggestion of the northern ring
around Westerlund~2, but it is not as clear as the one to the south,
mainly because the western side is absent as a result of the
``blister'' suggested by WU97.}
\label{fig:WU97-newreduction}%
\end{figure*}

The extended TeV source HESS~J1023$-$575 is positionally coincident
with the complex RCW~49/Westerlund~2 (Aharonian et al. 2007) and
constitutes evidence for the presence of relativistic particles in the
field.  Follow-up H.E.S.S. observations (Abramowski et al. 2011) still
did not determine the nature of the TeV emission.  Fermi observations
resolved the GeV emission into two different sources: 2FGL J1022.7--5741 and
2FGL J1023.5--5749c (Nolan et al. 2012). The first of these two sources
is now believed to be related to the pulsar PSR~J1022$-$5746 (Dormody
et al. 2009), while the second one is not yet clearly associated with
any known source (see Figure~\ref{finder-chart}).
One potential source of high-energy photons is relativistic particles 
that may be evident in radio observations. However, no clear 
connection between the radio emission in
the region and the high-energy sources has yet been established. 

The latest comprehensive study of RCW~49 at radio frequencies was
published by
Whiteoak \& Uchida (1997) [WU97]. The authors observed the source with
the Molonglo Observatory Synthesis Telescope (MOST) at 843 MHz and
with the Australia Telescope Compact Array (ATCA)\footnote{The
  Australia Telescope Compact Array is funded by the Commonwealth of
  Australia for operation as a National Facility by CSIRO.} at 1.4 and
2.4 GHz, attaining angular resolutions of 44'' to 7''. WU97 described
the detected emission as two ring-like features or shells (see
Fig.~\ref{fig:WU97-newreduction}). The authors proposed that the
northern one, with a blister toward the west, was created by the rich
cluster Westerlund~2 and that the wind of the well-known WR star in
the field, WR 20b, could have built the southern ring. The area where
the two shells seem to overlap shows the brightest radio emission, and
is identified here as the bridge.

Molecular lines have been studied in the region by several authors 
(Ohama et al. 2010 and references therein) at arcmin
angular resolution. For instance, Furukawa et al.
(2009) pointed out that several clouds exist with local standard of rest 
(LSR) velocities between $-11$ and $+9$ and between 11 and 21 km s$^{-1}$.

Spitzer IRAC observations of RCW~49 showed that dust coexists 
with ionized gas, or is embedded in neutral gas mixed with ionized gas 
(Churchwell et al. 2004). The data also let Whitney et al. (2004) 
conclude that star formation is occuring at present, continuous and/or  
triggered by stellar winds and shocks of the Westerlund~2 stellar
population. Deep cm-wavelength radio images at
a similar angular resolution are needed to clarify
the association between dust and and ionized matter. 
They could also pinpoint protostellar objects and star-forming
regions.

Very recently, the capabilities of the main radio interferometers have 
been considerably extended. The receiver bandwiths, inversely 
proportional to the square root of the
attainable noise, have been enlarged by a factor of ten and more.
We have obtained high-sensitivity arcsec resolution radio observations 
using the upgraded facilities
to search for point-source emission and extended low surface brightness
emission from the entire RCW~49 field and surroundings. We aim to look
for a correlation between the radio, IR, optical, and high-energy
emission in the field and study the radiation regime.

The content of this paper is as follows. Section 2 describes the new
broad-bandwidth observations and the reduction process. In Section 3
we report the data analysis, including an analysis of the spectral
index. Section 4 presents additional findings on correlations with
emission at other frequency ranges. Section 5 contains a discussion
of the distance to the complex and the putative relation
between the radio emission and the high-energy sources. Section 6
presents our conclusions and ideas for future work.

\section{Radio observations and data reduction}

\subsection{Continuum data}

Radio observations toward RCW~49 were carried out with the Australia
Telescope Compact Array (ATCA, project C1847) 
using the Compact Array Broadband Backend
(CABB, Wilson et al. 2011). Two array configurations were used,
6A on 2012 February
20 and 750D on 2012 February 22, for 12 hours each. 
Observations at 5.5 GHz (C band) and 9.0 GHz (X band) were obtained 
simultaneously, each with a bandwidth of 2 GHz.
They were planned to derive
information on the spectral index of the radiation. Hence, 5.5 and 9.0
GHz were chosen for their high resolution (relative to 1.4 and 2.4
GHz) and as frequencies where any potential nonthermal emission can
still be detected. With the combination of configurations
and frequencies mentioned above we expected to be able to image structures
between 1 and $\sim$100'' in extent.

The observed field comprised the extension of the TeV source HESS
J1023--575 and the cm-radio emission measured by WU97, covering an area
of $\sim$ 30 arcmin$^2$. The observing strategy consisted of building
a mosaic with 41 pointings, ensuring Nyquist sampling at the higher
frequency band. The integration time for each pointing was 
approximately 13 minutes. 
The bright source PMN J1047-6217 was used to calibrate the
antenna gain phase, and monitored before and after each mosaic
observation so that the gain phase could be interpolated through the
mosaic observations. The absolute flux scale was determined from PKS
1934-638, assuming flux density values of $5.1$ Jy (5.5 GHz) and
$2.74$ Jy (9.0 GHz).

   \begin{figure*}[!t]
   \centering
  \includegraphics[width=0.78\textwidth,angle=0]{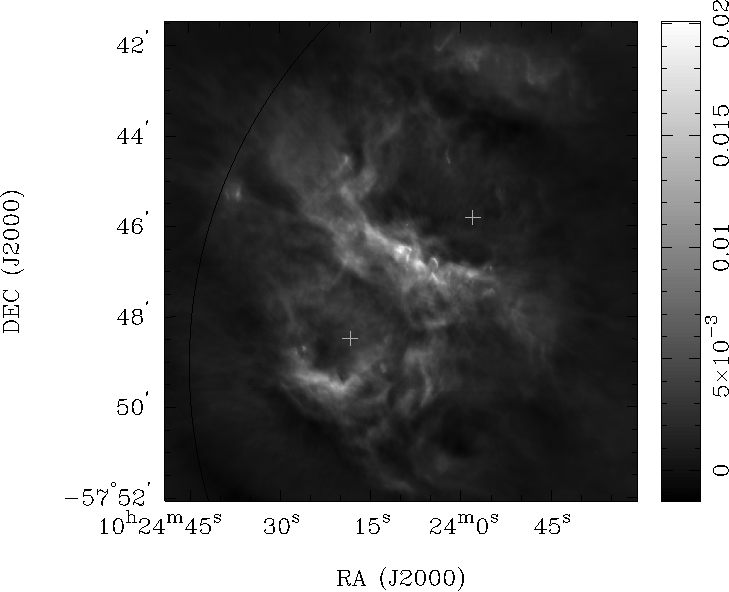}
  \includegraphics[width=0.73\textwidth,angle=0]{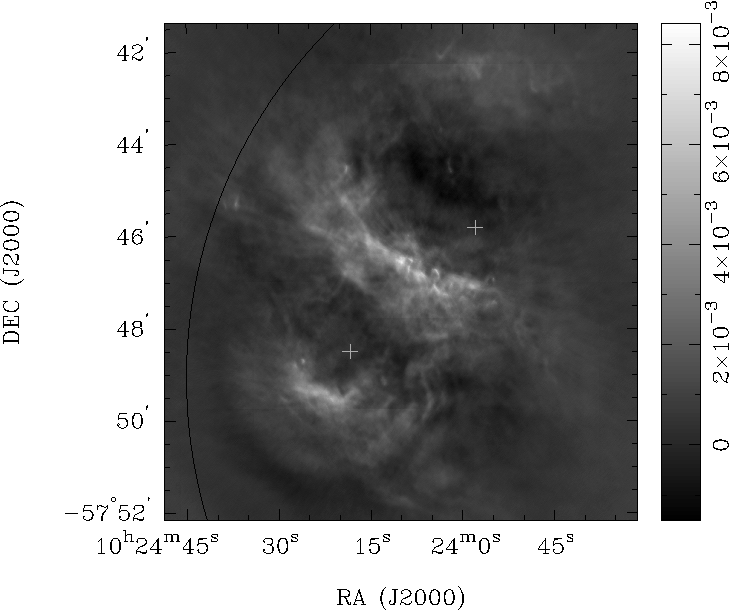}
   \caption{{\sl Top:} The 5.5-GHz ATCA image of RCW~49 field. The
     synthesized beam is $1.9''\times 1.5''$.  {\sl Bottom:} The
     9.0-GHz image of the RCW~49 field. The synthesized beam is
     $1.2''\times 0.9''$. The brightness level in Jy beam$^{-1}$ is
     shown to the right of each image. The position of the stars 
     WR 20a (N) and WR 20b (S) is indicated with light-gray crosses.
     The dark-gray circular sector represents the angular size of the 
     extended TeV source HESS~J1023$-$575.}
   \label{fig:CABBgrayscale}%
    \end{figure*}

%
\begin{table}[h]
\caption{Main parameters of the CABB ATCA images at 5.5 and 9 GHz.}
\label{table1}    
\centering         
\begin{tabular}{l r r}
\hline\hline     
Parameter    & 5.5-GHz image & 9.0-GHz image \\    
\hline       
 Synthesized beam & $3''\times 3''$ & $2''\times 2''$\\
 rms (mJy beam$^{-1}$)& 0.6 & 0.6 \\
 Min/Max (mJy beam$^{-1}$) &  -1.3/22.6 & -1.4/55.9\\ 
 Flux density$^\dag$  (Jy) & 146$\pm$12 & 103$\pm$30  \\
 Baseline uv range (k$\lambda$) & 0.5 -- 129 & 0.8 -- 198 \\
\hline                 
\end{tabular}

\tablefoot{$\dag$: Total flux density measured above 5$\sigma$ (=5 rms) 
over the RCW~49 region.}
\end{table}

The multichannel image reconstruction, image analysis and display
({\sc miriad}) 
routines\footnote{www.atnf.csiro.au/computing/software/miriad/} (Sault
et al. 1995) were used to
perform the data editing, calibration and image reconstruction.  A
number of factors complicated the reduction and imaging process,
namely: {\sl (i)} the large angular extension of the radio source,
which required mosaicing; {\sl (ii)} the complex nature of the radio
source with a highly dynamical range of the emission comprising bright
ridges, faint extensions, loops, and shells; {\sl (iii)} the broad CABB
bandwidth of 2-GHz presents a high fractional bandwidth and
correspondingly wide ranges in primary and synthesized beams, plus
variations in the emission level across the band due to spectral index
variations; and {\sl (iv)} the ability of {\sc miriad} tasks to deal
with these demanding observations and large data sets.

Various approaches were tested to construct the images using the tasks
{\sc MOSSDI, MOSMEM}, and {\sc MFCLEAN}. It was found that the maximum entropy
deconvolution {\sc MOSMEM} was the best tool not only for minimizing side lobes,
but for dealing with different levels of radio luminosity from the various
pointings. Results from the {\sc CLEAN} algorithm were also examined, but
remained less satisfactory, as expected for extended
low-surface brightness emission when compared with maximum entropy
methods. A similar rms noise level of $\leq$ 0.6 mJy beam$^{-1}$ was
attained at both frequency bands, as determined from outer regions of
the field where no emission from RCW~49 was evident. This noise level
is equivalent to $3\times$ and $2\times$ the theoretical rms values at
5.5 and 9 GHz, respectively. The resulting images at the two
frequencies are shown in Fig.~\ref{fig:CABBgrayscale}.

Goss \& Shaver (1970) imaged the RCW~49 complex with the Parkes
radiotelescope at 5000 MHz (FWHM: 4.1', see their Fig. 6). The
authors detected the source G284.3$-$03, positionally coincident
with RCW~49, as well as a much fainter source to the east (G284.6$-$0.2),
and quoted an integrated flux for both sources of 335 Jy.
Later on, Churchwell et al. (1974) observed the region in continuum and
line emission. They described the region as an area of $5'\times 7'$
and derived an integrated flux density of 178.8 Jy at 5 GHz.
The integrated flux at 5.5 GHz for the data presented here, above 
5$\sigma$, is given in Table 1, and above 3$\sigma$ is $\sim 160\pm15$ Jy.
A comparison between this last flux density value and that published by
Churchwell et al. (1974) shows that the interferometric data gather
about 90\% of the flux measured by a single-dish telescope.\\

\subsection{Line data}

We took advantage of the CABB-ATCA simultaneous zoom modes and
performed a pilot line observation with the 750-D configuration.  The
central frequency was 5005 MHz, at which two transitions are expected:
the H$137\beta$ radio recombination line and the $3_1 - 3_1$ A-branch
line of methanol (Robinson et al. 1974). 
We set the zoom modes of the correlator to cover 2560
channels of 0.5 kHz width each. This corresponds to a velocity
resolution of 0.05 km s$^{-1}$. The final line-only data cube was
built with robust weighting set to zero, applying hanning smoothing to
reduce the number of channels to 270, each of 0.5 km s$^{-1}$ width.

\subsection{Archive 1.4 and 2.4 GHz data}
\label{subsec:archivedata}
The archive 1.4 and 2.4 GHz image data sets from the WU97 project,
campaign C492, were re-reduced and imaged. The project consisted of a
12-h observation at 1.5C array and a 12-h observation at the 1.5B
array, with central frequencies of 1.376 and 2.378 MHz, 128 MHz
bandwidth each. As with the new observations presented in this paper,
the archival data were edited and calibrated with the {\sc miriad}
package. Images were built using maximum-entropy algorithms with a
synthesized beam of 10 arcsec. The resulting image rms at 1.376 GHz
was 3 mJy/beam, and at 2.378 GHz it was 1 mJy/beam.

\section{Analysis of the radio data}

The ATCA 5.5 and 9 GHz continuum images (Fig.~\ref{fig:CABBgrayscale})
show a high degree of detail compared with the observations of WU97
(their Fig. 3, and Fig.~\ref{fig:WU97-newreduction} presented here),
which were described by the northern and southern shells, separated by
the bridge. Figure~\ref{fig:CABBgrayscale} confirms this simple
picture for the largest-scale emission in RCW~49, but at the highest
resolution the new data reveal much more complex structures.  At both
frequencies, complexes of pillars or fingers of emission are very
clear in the bridge. A plethora of filaments of different sizes,
widths, and intensities dominate the emission. Clouds and
circular-elliptical sources of a few seconds of arc are also visible.
Though somewhat less evident, there is at least one bow-shock-like
feature at RA, Dec (J2000)= 10:24:38.73, $-$57:45:19.9, marked in
Fig.~\ref{fig:discr-scs} as S3.

There are emission minima in the vicinity of the stars WR~20a 
and WR~20b. The average minimum flux density in the
region of Westerlund~2 is $\sim$1 mJy beam$^{-1}$ at 5.5 GHz, and
$\sim-0.5$ mJy beam$^{-1}$ at 9 GHz. The corresponding one in the
region of WR 20b is $\sim$0.7 mJy beam$^{-1}$ at 5.5 GHz, and
$\sim$0.3 mJy beam$^{-1}$ at 9 GHz. 

The observed field contains some of the brightest massive, early-type
stars in the Galaxy. No radio emission from the direction of the
3.7-day period massive binary WR 20a, WR 20b or the Westerlund~2 core
is detected above 5-$\sigma$ level in the images shown in
Fig.~\ref{fig:CABBgrayscale}. A search of an image generated only with
high-spatial frequency data, that is, long baselines, to reduce the
impact of the large-scale diffuse emission also failed to detect any
point source emission in these directions. We do not expect to detect
thermal wind sources.  For a canonical WR star with a mass-loss rate of
$~\sim3\times10^{-5}$~M$_\odot$~yr$^{-1}$ (e.g. Chapman et al. 1999)
and wind speed $\sim1000$~km\,s$^{-1}$, the expected thermal flux is
0.3~mJy for a distance of 4~kpc.

\subsection{Discrete sources in the surroundings of RCW~49}

\begin{figure}
   \centering
  \includegraphics[width=0.5\textwidth,angle=0]{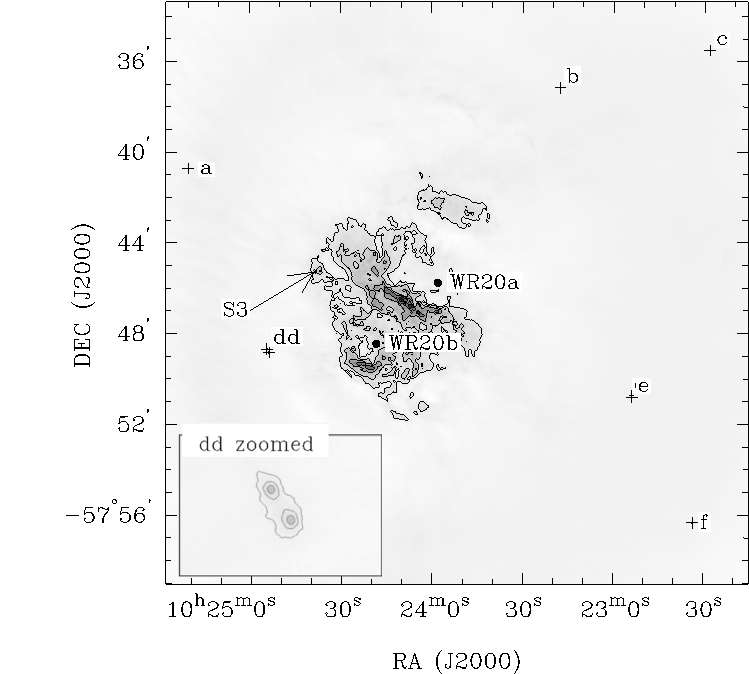}
   \caption{ATCA 5.5 GHz image of RCW~49 and surroundings. 
     The contour levels are -0.6, 3, 9, 20, 35, and 60 mJy beam$^{-1}$; 
     the synthesized beam is $3''\times 3''$. Crosses mark the positions
     of radio sources detected near RCW~49; these are cataloged in
     Table~\ref{tab:disc-tab}. The arrow shows the 
     position of a bow-shaped structure, S3 (see Sect. 3.4). 
     The WR stars 20a and 20b are marked with solid circles.
     The inset shows the emission of source 'dd', with contour levels of 2, 4, 
     and 8 mJy beam$^{-1}$.} 
   \label{fig:discr-scs}%
    \end{figure}

We identified six discrete isolated radio sources in the 5.5-GHz continuum 
image in field surrounding RCW~49 labeled 
'a', 'b', 'c', 'dd' (double), 'e', and 'f' (see
Fig.~\ref{fig:discr-scs}).

In Table 2 we list the results of Gaussian fits to the discrete
sources: peak coordinates (J2000), total flux density, and peak flux
density. The Simbad and NED databases were searched for
counterparts at these positions. We also searched the DSS2 and
Spitzer-GLIMPSE images. The results of the search are also listed in
Table 2, where $D$ is the angular distance between the discrete source
and the nearest cataloged object. Given the separations of potential
counterparts from the radio source positions relative to the 2-arcsec
beam of the radio data, we conclude that we only detected radio emission
from one previously identified source, dd1.

The double source dd, composed of dd1 and dd2, was also detected at
9.0 GHz. The proposed counterpart, positionally coincident with dd1,
was identified by Tsujimoto et al. (2007) as a young stellar object
(YSO) candidate. Both radio components are discrete sources at 5.5 and 9
GHz resolution, with total flux densities at 9.0 GHz of $S({\rm dd1})
= 12.2$ mJy and $S({\rm dd2}) = 12.1$ mJy. The fitted rms at 9.0 GHz
is 0.1 mJy beam$^{-1}$. The spectral index between 5.5 and 9 GHz of
each component is $\alpha ({\rm dd1}) \sim -0.9$ and $\alpha ({\rm
  dd2}) \sim -0.8$.  Only the component dd1 
appears to correspond with the position of the YSO candidate.
However, taking into account the morphology and the spectral 
indices, we propose that the source dd probably has an extragalactic 
origin\footnote{The number of expected extragalactic 
sources in the images presented here
can be estimated. Following either Anglada et al. (1998) or,
independently, the results previously obtained by Purcell et al. (2010) of the
CORNISH project, the expected extragalactic background objects, 
above 5 $\sigma$ (=3 mJy) in a $(10')^2$ (=0.03 sq deg) field area, at the
5.5-GHz band, is close to unity.}.

\begin{table*}[t]
\caption{Parameters of discrete 5.5 GHz radio sources detected toward RCW~49.}
\label{tab:disc-tab}    
\centering         
\begin{tabular}{l l r r l l l}
\hline     
Id & RA,Dec (J2000)  & Total $S_{\rm 5.5GHz}$ &  Peak $S_{\rm 5.5GHz}$ & 
Nearest cataloged source & $D$ & Comments \\    
     & (hms, dms)     &       (mJy)  & (mJy/beam)  & & (arcsec) & \\
\hline       
a & 10:25:20.40, --57:40:43.3 & 12$\pm$2 &  3.4$\pm$0.6 & IRAS 10236--5723 & 115 & \\
b & 10:23:17.70, --57:36:12.4 & 23$\pm$2 &  1.8$\pm$0.1 & 2E 1021.5-5720 & 50 & X-ray source$^1$\\
& & & & 1RXS J102322.2--573548 & 48 & ROSAT source$^2$ \\
& & & & CXO J102323.3--573749 & 108 & Chandra source$^3$ \\
c & 10:22:28.36, --57:35:30.9 & 3.5$\pm$0.5  &  1.6$\pm$0.1& TYC 8608--1017-1 & 62 & Star \\
dd1 &10:24:54.61, --57:48:42.9 & 12$\pm$1 & 10.7$\pm$0.2 & CXOU J102454.5-574842 & 0.24 & YSO candidate$^4$  \\
dd2 & 10:24:53.89, --57:48:52.2& 14$\pm$2 & 10$\pm$0.8 & CXOU J102454.5-574842 & 12 & YSO candidate$^4$\\
e & 10:22:53.82, --57:50:46.1 & 7$\pm$1  & 3.2$\pm$0.4 & TYC 8608--1463--1 & 40 & Star \\
f & 10:22:33.58, --57:56:20.9 & 15$\pm$2 & 15.3$\pm$0.8 &TYC 8608--366--1 & 21 & Star\\
\hline                 
\end{tabular}
\tablefoot{1: Belloni and Mereghetti 1994, 2: Voges et al. 2000,
3: Evans et al. 2010, 4: Tsujimoto et al. 2007.}
\end{table*}


\subsection{Spectral index analysis}
\label{sec:spix}

Spectral index maps
were generated between pairs of observation data sets taken
simultaneously with the ATCA.

For each set of data in the analysis, we prepared the data at the two
frequencies in the following way: we considered visibilities from
baselines sensitive to emission at the same angular scales, that is,
we limited the $uv$ range to the same lower limit at both
frequencies. We also convolved the higher-frequency map with a
synthesized beam of the same size as that of the lower frequency
data. In assembling the spectral index maps and spectral index error
maps we used only pixels with a signal-to-noise ratio higher than 5. 
It should be
noted that this analysis does not include the flux on the largest
scales, that is, at the lowest spatial frequencies, since they are not
sampled at both observing frequencies. Thus, the spectral index
information must be interpreted with some caution.

\begin{figure}
   \centering
   \includegraphics[angle=-90,width=0.47\textwidth]{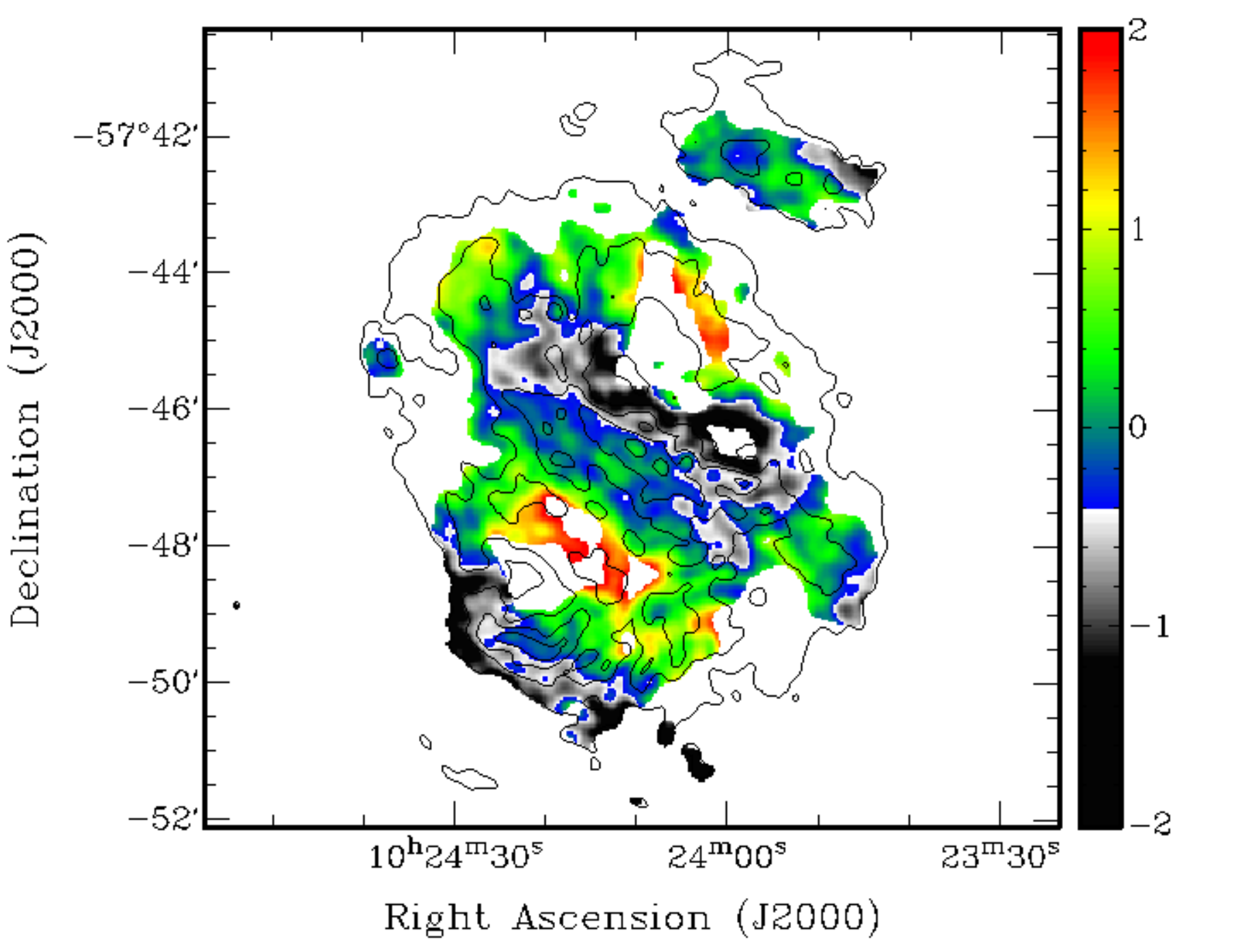}
   \includegraphics[angle=-90,width=0.47\textwidth]{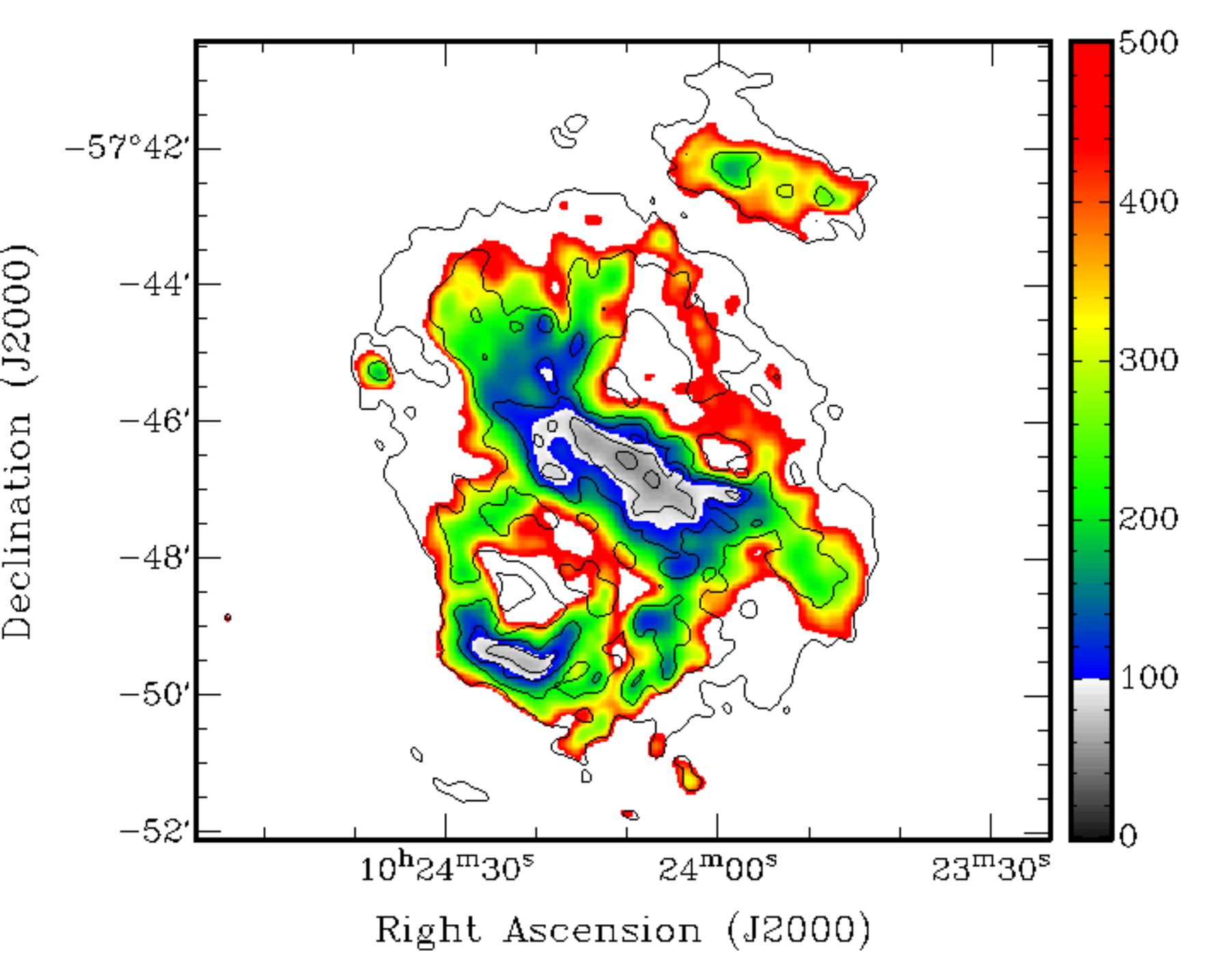}
  \caption{{\sl Top:} Spectral index distribution image using the WU97
    1.4-GHz and 2.4-GHz data at baselines $\geq$ 0.25 k$\lambda$.  
    This value corresponds to angular scales of $\sim$ 25'.
    The black contours represent continuum emission at 2.4 GHz.  {\sl
      Bottom:} Spectral index error distribution for the spectral
    index values above, in thousandths.}
   \label{fig:spixwu}
    \end{figure}

\paragraph{\bf Map between 1.4 and 2.4 GHz pre-CABB ATCA data.} The data sets 
at 1.4 and 2.4 GHz (project C492, Whiteoak \& Uchida 1997) comprised
observations performed simultaneously at the two bands. 
We restricted the generation of the images to visibilities with a
minimum $uv$ range of 0.25 k$\lambda$.  Figure~\ref{fig:spixwu} presents
the resulting spectral index map, along with the uncertainty map.

The majority of the region where the 1.4-GHz emission is brightest
(the bridge) has spectral index values $\sim0$, with uncertainties of
less than 0.2. We interpret this as optically thin thermal emission
that we expect in a large diffuse HII region such as RCW~49. However,
we also see some evidence in the bridge of spectral index values
that are steeper than $-$0.4, with relatively low uncertainties.  Of
most interest are the areas steeper than $-$0.4 that are in the areas
with some of the brightest 2.4-GHz emission. Outside these bright
regions, the indication of steep indices is less robust since we may
be missing some emission at the higher frequency on the larger scales,
as inferred by the steepest indices appearing where the high frequency
emission level is low, that is, on the northern edge of the bridge and the
southern edge of the south shell.

\begin{figure}
   \centering
   \includegraphics[angle=-90,width=0.47\textwidth]{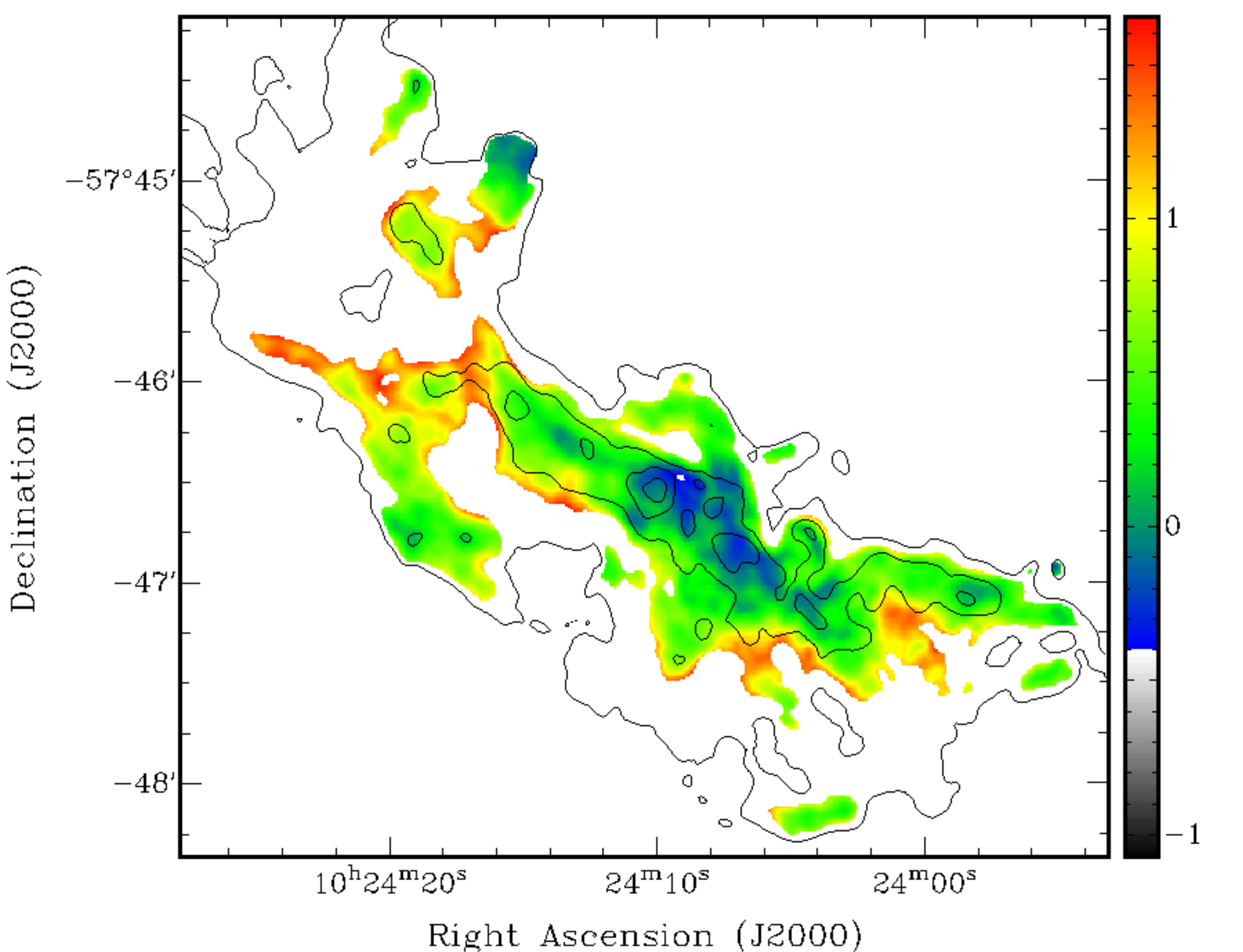}
   \includegraphics[angle=-90,width=0.47\textwidth]{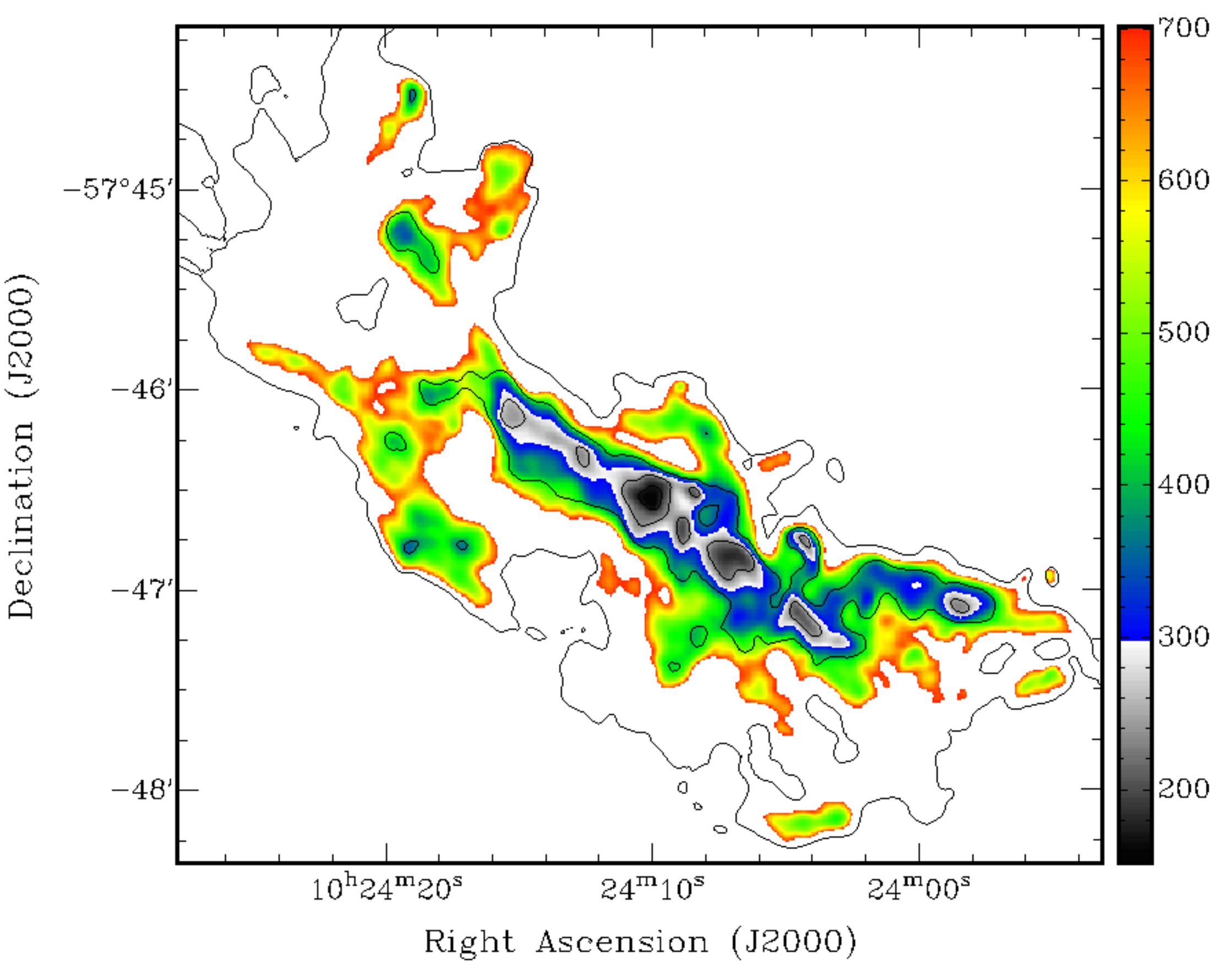}
  \caption{{\sl Top:} Spectral index distribution based on 5.5 and
    9.0-GHz for baselines $\geq$ 1k$\lambda$. Contours represent the
    9.0-GHz continuum data. {\sl Bottom:} Spectral index error
    distribution image, in thousandths.}
   \label{fig:spix5-9}%
    \end{figure}

\paragraph{\bf Map between 5.5 and 9.0 GHz CABB data.} In configurations 
6A+750D, the $uv$ range covered at 5.5 GHz by the 2-GHz bandwidth is
(0.5, 129) k$\lambda$ (Table 1). At 9.0 GHz, the baseline range is (0.8,
198) k$\lambda$. We built spectral index images using the same minimum
baseline value of $\sim$ 1k$\lambda$. Figure~\ref{fig:spix5-9} shows the
resulting spectral index map and the spectral index error
distribution.

The bridge is the only part of the region that has sufficient signal
for spectral index values to be determined. As in the 1.4 and 2.4 GHz
index map, the bulk of the emission in the bridge has index values
$\sim0$, but there appear to be some areas, associated with the 
brightest emission, that have values $\sim-0.4$, with low uncertainty.

\paragraph{\bf The 2-GHz bandwidth at 5.5 GHz.} 
The large 2-GHz extent of the CABB bandwidth allowed us to try the
experiment of dividing the 5.5 GHz data into two data sets, with
central frequencies of 5.02 and 5.94 GHz, both with a bandwidth of 460
MHz.  The continuum images at 5.02 and 5.94 GHz revealed different
flux levels.  However, the spectral index map built using these data
had large uncertainties due to the frequency proximity of the two
data sets, and we abandoned any idea of interpreting the data.

\subsection{Line results}

\begin{figure}
\centering
\includegraphics[width=0.5\textwidth,angle=0]{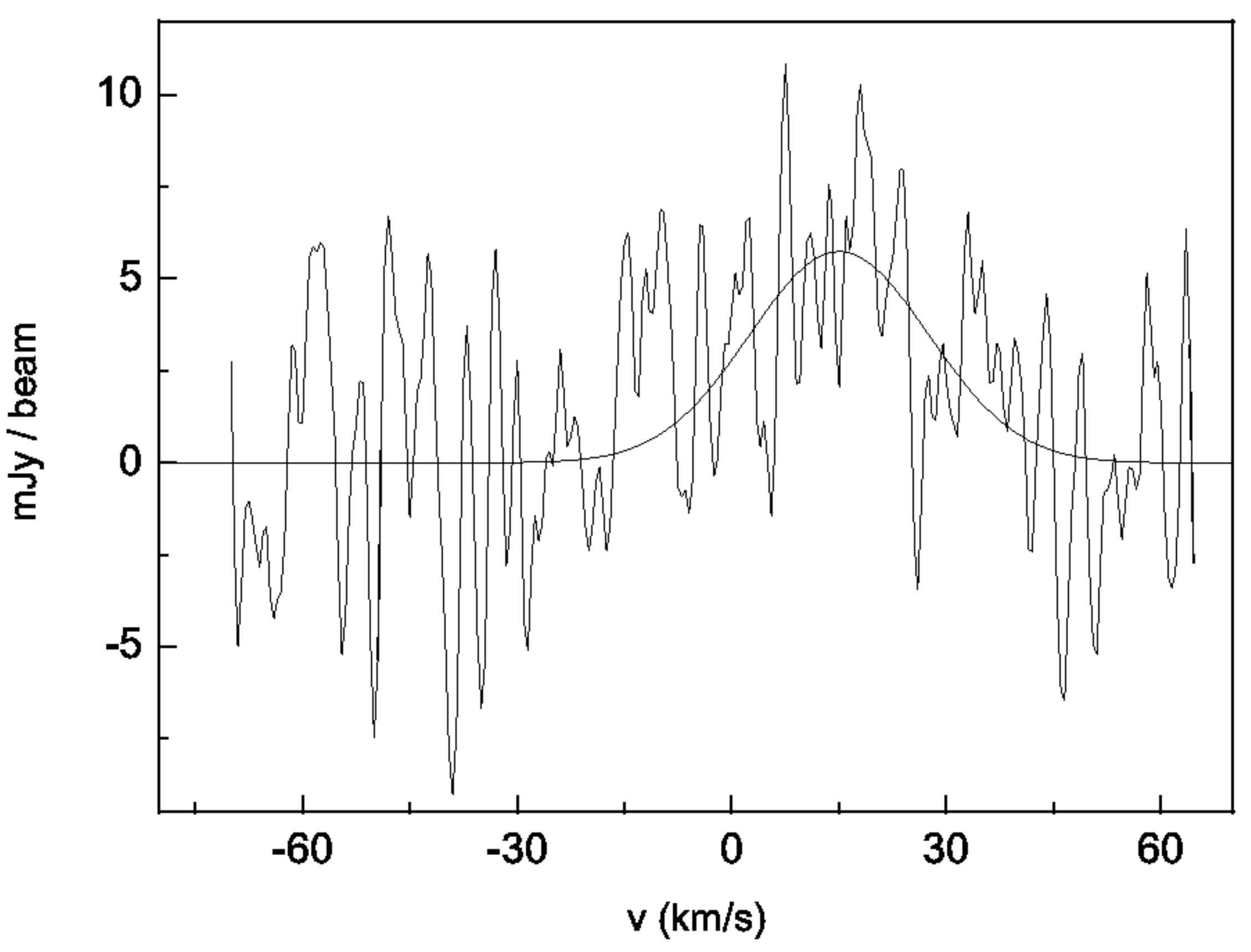}
  \caption{Spectrum of the RRL H$137\beta$ line near 5005 MHz,
  averaged over the maxima of the 5.5 GHz ATCA continuum image
  ($S_{\rm cont} \geq 30$ mJy beam$^{-1}$, and Gaussian fit with a peak
  flux of 6 mJy beam$^{-1}$, a central velocity (LSR) of +15 km s$^{-1}$,
  and a velocity width of 25 km s$^{-1}$).}
   \label{rrline}%
    \end{figure}

Figure~\ref{rrline} 
shows the spectrum at 5.005 GHz averaged over the continuum emission
bridge region, over pixels with $S > 30$ mJy beam$^{-1}$ in the
radio continuum. There is
emission detected above the noise level at the frequency of the RRL
H137$\beta$ transition.

The expected rms
that corresponds to the instrumental settings we used\footnote{www.narrabri.atnfcsiro.au/myatca/sensitivity\_calculator.html}
is 1.8 mJy beam$^{-1}$. The measured rms is 2 mJy  beam$^{-1}$.
The detected line has a peak flux density of $\sim$ 6 mJy at a center
velocity of 15 km s$^{-1}$ and a velocity width of 25 km s$^{-1}$.
The continuum flux density, averaged in the same region, is $S_{\rm c}
=2.5$ Jy. 

Caswell \&  Haynes (1987) listed RRL emission from RCW~49 at the 
H109$\alpha$ line with a central velocity of 0 km s$^{-1}$ and a
velocity width of 46 km s$^{-1}$. Churchwell et al. (1974) also
detected 
He109$\alpha$, with a central velocity of $-4\pm1$ km s$^{-1}$, and 
a velocity width of about 50 km s$^{-1}$ (data collected with the
single-dish Parkes telescope). 

In search for anomalous microwave emission
on small angular scales of the RCW~49 core, 
Paladini et al. (2013) observed the bridge region 
at H109$\alpha$ with the ATCA (in 2009). They drew spectra at four positions
along the bridge and detected the line at three of them. Their measured 
central velocities correspond to
14.5$\pm$1.8, $-4.1\pm$5.5, and 14.5$\pm$1.8 km s$^{-1}$,
in full agreement with the parameters found here for H137$\beta$.
These velocities are also consistent with those of the highest
velocity molecular cloud reported by Furukawa et al. (2009).

   \begin{figure}[!t]
  \centering
  \includegraphics[width=0.45\textwidth,angle=0]{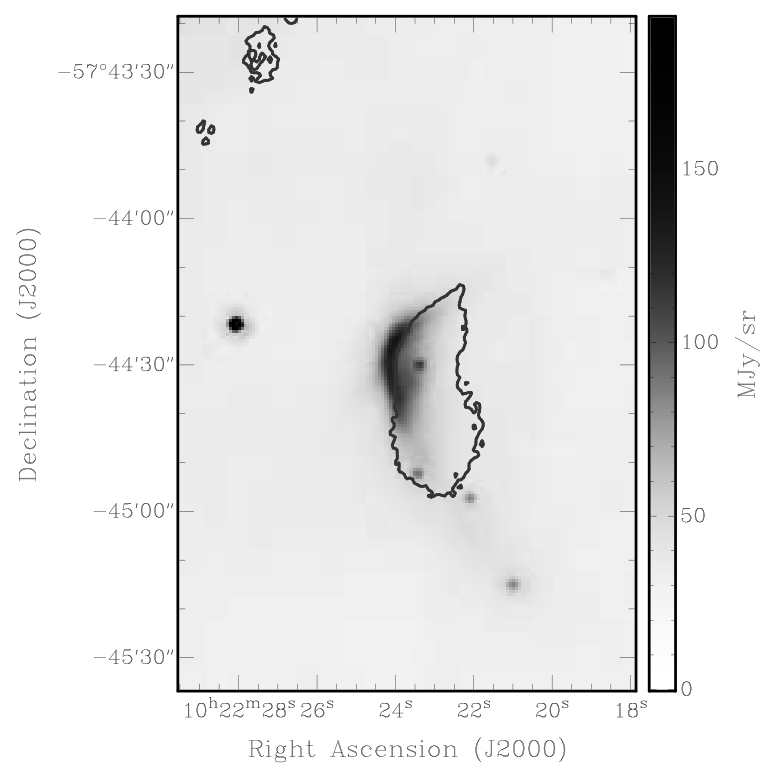}
 \includegraphics[width=0.45\textwidth,angle=0]{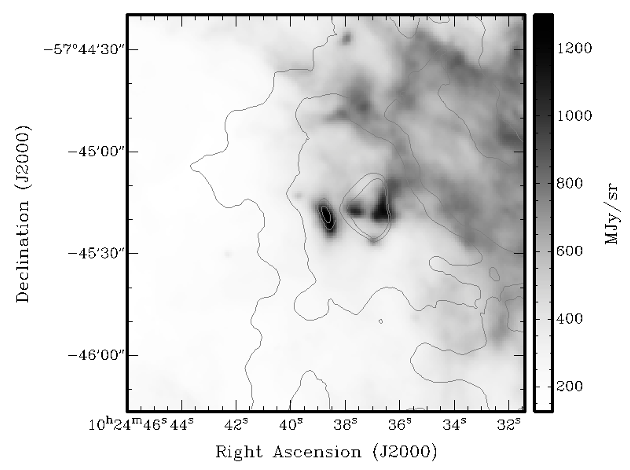}
   \caption{{\sl Top:} ATCA 5.5 GHz continuum emission contours
     toward the object S1 from Povich et al. (2008). Levels
     are at 1.8 and
     3 mJy beam$^{-1}$. In grayscale: emission at 8$\mu$m from
     Spitzer-IRAC.  {\sl Bottom:} 5.5-GHz emission toward the object
     S3 from Povich et al. (2008) at the center of the image. Levels are
     at 3, 5, 11, and 12 mJy beam$^{-1}$. In grayscale: emission at 8$\mu$m
     from Spitzer-IRAC.}
   \label{fig:s1-s3}
    \end{figure}

\subsection{Stellar bow shocks}

Using Spitzer-GLIMPSE data, Povich et al. (2008) discovered three
stellar bow shocks in the region of RCW~49 and named them RCW~49-S1,
-S2, and -S3. The authors proposed that the regions are physically 
related to RCW~49
and that S2 and S3 still were inside the HII region. They
suggested that S1 may be formed by the combined action of the stellar
winds of Westerlund~2 stars that have escaped from the HII region at
hundreds of km~s$^{-1}$.

In addition to the detection at the GLIMPSE bands, RCW~49-S1 is a point
source at MSX and IRAS images. The exciting star would be an O5 III if
the distance to the complex is 6.1 kpc (Povich et al. 2008).  The
spectral energy distribution of source S1 has been modeled by Povich
et al. (2008, see their Fig. 3: Top). To match the 60 and 100 $\mu$m
IRAS fluxes 
the authors assumed the presence of low-density material farther
from the star. They produced a model of the emission from a shell that
is 2 -- 3~pc from the star. Their model predicts a shell flux of $\sim$
0.15 Jy at 1000 $\mu$m.

We searched for ATCA-5.5 GHz continuum sources at the positions of the
bow shocks S1, S2, and S3. The images revealed emission consistent with
bow-shock features positionally coincident with RCW~49-S1 and
-S3. Even if S2 radiates in radio, intense extended 5.5 GHz emission
present at the position of S2 will be hiding the feature from view.

Figure~\ref{fig:s1-s3} shows the 5.5 GHz continuum emission at the
locations of S1 and S3.  The total flux densities are 70$\pm$10 mJy
and 140$\pm$20 mJy, respectively.  There is some emission at
the position of S3 at 9 GHz but it is below 3$\sigma$.
S1 is off the 9-GHz continuum mosaic image.

Assuming that the gas forming the radio source coincident with S1 
is optically thin ($\alpha \sim 0$),
the expected flux at 1000 $\mu$m is $S_{\rm 1000 \mu m} \approx$ 0.07 mJy. 
If the 5.5 GHz source is a bow shock, one can use the $S_{\rm 1000 \mu m}$ value
as a key parameter 
to distinguish between different models of envelopes developed as
in Povich et al. (2008).

\subsection{Surroundings of star MSP 18}

\begin{figure}
  \centering
  \includegraphics[angle=0,width=0.45\textwidth]{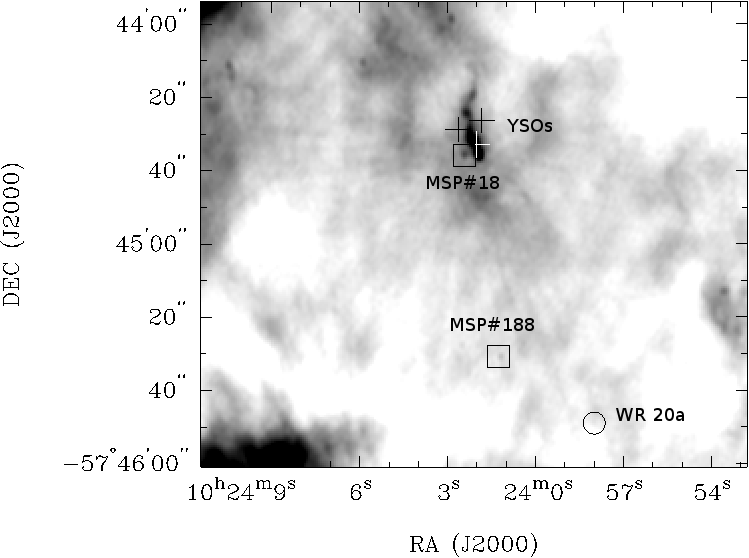}
  \includegraphics[angle=0,width=0.45\textwidth]{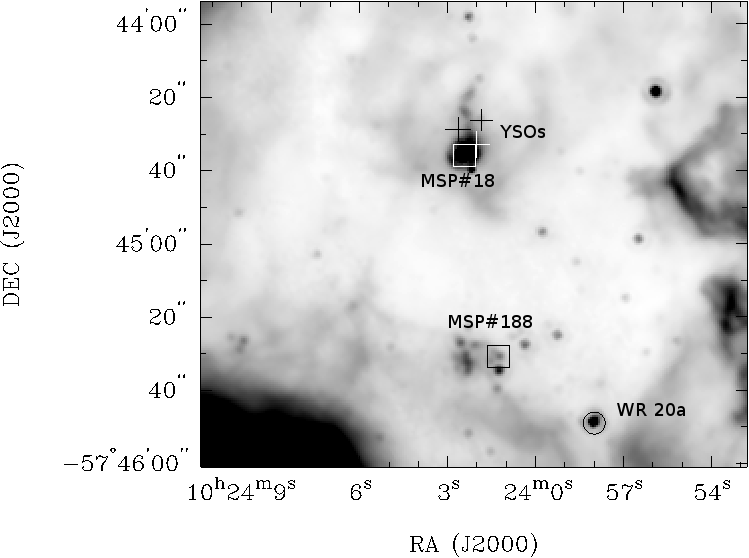}
   \caption{{\sl Top:} Radio emission at 5.5 GHz showing a jet-like
     source.  {\sl Bottom:} The same region at the 8$\mu$m-IRAC
     band. Surrounding the jet-like source appears a closely circular
     ring of emission. The symbols represent the positions of known YSO
     candidates (crosses) and the Westerlund~2 stars MSP\#18, MSP\#188 
     (open squares), and WR 20a (circle).}
  \label{fig:jet}%
   \end{figure}

To the north of the region of Westerlund~2 we have discovered a
jet-like structure at both radio bands (Fig.~\ref{fig:jet}),
of $\sim$ $20''$ (0.01 pc if the distance is 6 kpc).
There are
a number of known sources closely associated with the position of the
jet, including two YSO candidates, a pre-main-sequence star, and the
bright O-supergiant MSP \#18.  This last object is a Westerlund~2
member, marginally detected in the 5.5 GHz full angular resolution
image. The peak flux is 3.1 mJy beam$^{-1}$, and the background level
at its position is $\sim$ 2.5 mJy beam$^{-1}$ (see
Sec.~\ref{sec:distance} for a discussion).

It is not clear if any one of the objects mentioned before is the
source of the jet. The jet-like structure is surrounded by filaments
at both sides, which appear to be forming a shell or bubble.

We estimated the spectral index value of the brightest part of the jet
to be $-2.2\pm0.3$, based on an average taken over the pixels where
the spectral index error was below or equal to 0.3. Such a negative spectral
index denotes the conjunction of a steep injection spetrum in addition to very
strong cooling ($\propto E^2$, Vila \& Aharonian 2009, p. 24).

The jet-like source is also detected in the 8 $\mu$m-IRAC band
(Fig.~\ref{fig:jet}, bottom).  An investigation of the nature of the
jet-like structure, its possible physical association to the bubble,
energetics, and the relation with other sources in the field is under way
and will be reported elsewhere.

\section{Radio, IR, and X-ray emission from RCW~49}

\begin{figure*}[!t]
  \centering
\includegraphics[angle=0,width=0.8\textwidth]{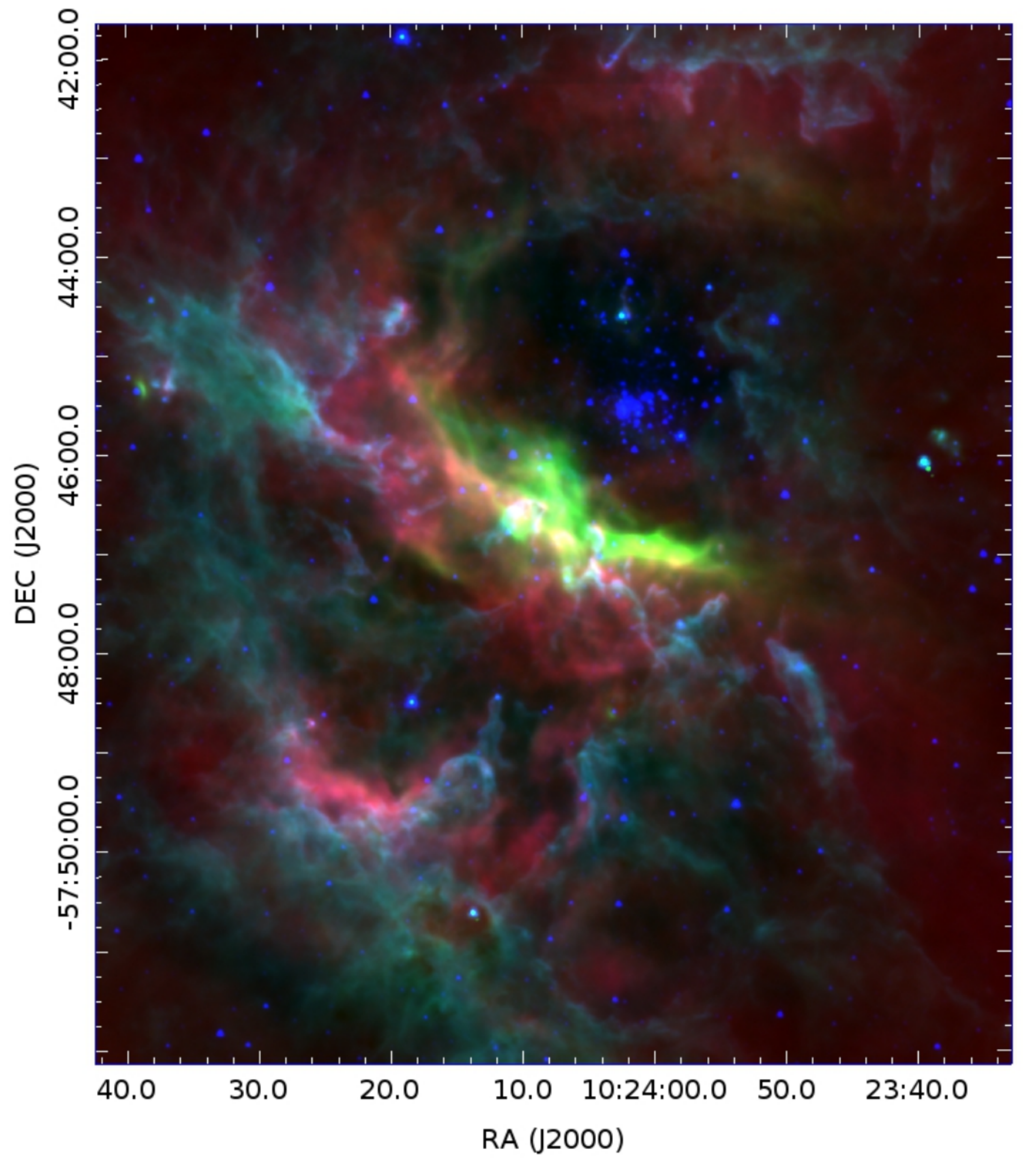}
   \caption{RCW~49 field as seen at radio continuum (9 GHz, in
red) and Spitzer-GLIMPSE band 1 (3.6 $\mu$m) in blue and band 4
(8 $\mu$m) in green. The stars in the field of view, including the core
of the cluster, clearly visible at 3.6 $\mu$m. The close association of the near-IR
and radio emission is evident in the north and south shells and in the bridge. }
   \label{fig:irradio}%
   \end{figure*}

Fig.~\ref{fig:irradio} portrays the emission at radio and near-IR,
showing the free-free emission from the ionized gas 
and dust (polycyclic aromatic hydrocarbons, PAH) emission.  
The stars from the Westerlund~2
cluster are detected at IRAC band 1. There is some intense band 4
emission (8 $\mu$m) in the bridge region.
At the frequency of IRAC-band 4 the emission from PAHs is higher 
than at other bands (see Reich et al. 2006, their Fig. 1). 
Bright emission accounts for a dust-rich region, with the consequent
effect of radiation reddening.
IRAC band 4 also shows a discrete source at the position of
the star MSP \#18, near the jet-like source described above.

The structures revealed in the radio are very similar to those
observed in the near-IR by GLIMPSE. The northern and southern shells,
and the bridge described by WU97 are very evident. It is clear that
the HII region is not in an area of uniform density, with a lack of
near-IR emission on the west side of RCW~49 suggesting lower material
density to the west. This explains the possible source of the radio
``blister'' structure on the west side of the RCW~49 H II region
mentioned by WU97.

We correlated the 5.5-GHz continuum data with Chandra data 
taken by Naz\'e et al. (2008) and found no apparent associations.

\paragraph{\bf The Fermi-pulsar region.} We surveyed the region of the 
Fermi pulsar 2FGL J1022.7--5741, searching for a signature of a pulsar
wind nebula. There is some emission at a level of 2 mJy, but it
remains close to the noise. A higher-sensitivity observation is needed
to image this emission and study weather it is associated with the pulsar.

\section{Discussion}

\subsection{Distance to RCW~49 and Westerlund~2}
\label{sec:distance} 
The distance to Westerlund~2 and RCW~49 is somewhat uncertain owing
to the large scatter in the values derived (see
Table~\ref{tabdistance}). The bulk of the distance estimates falls
between 5 and 8 kpc, but there are some notable outliers. The radio
observations presented here let us draw some general conclusions
related to the distance discussion.

\paragraph{\bf The HII region RCW~49.} The central velocity of the putative 
H$137\beta$ line detected in this study is similar to one of the CO
line components found by Furukawa et al. (2009) and Ohama et
al. (2010), to the 'B' component in the spectrum shown by
Dame (2007), and to the H109$\alpha$ line recently found by
Paladini et al (2013). The central line velocity 
of +15km/s corresponds to a
kinematic distance of $\sim6$~kpc, according to the rotation curve in
the outer galaxy derived by Brand \& Blitz (1993). The emitting
ionized gas corresponds to the brightest continuum features that form
the RCW~49 region. The fact that the ionized gas and part of the CO
gas have similar velocities suggests that the ionized and the
molecular components are physically associated.

\paragraph{\bf The cluster Westerlund~2.}  Even if Westerlund~2 is 
located 
as close as 2.5 kpc, the thermal flux from the individual stellar
winds of cluster members would 
not be detectable, well below the noise level of
the images presented here. We estimate the expected 5.5 GHz flux from
the the late O-type supergiants in the cluster (e.g. Moffat et
al. 1991) as 0.1 mJy for a distance of 2.5 kpc, and as 0.01 mJy 
for a distance of 6 kpc.

At the position of the Westerlund~2, the 5.5-GHz continuum emission
shows diffuse emission at a level of $\sim3$~mJy. No correlation
between stellar positions and peaks in the radio flux were found,
except for maxima in the radio emission of 1.6 mJy beam$^{-1}$ at the
exact position of MSP 188, and of 3.1 mJy beam$^{-1}$ at the location
of MSP 18. Given the 3.0~mJy of diffuse emission, these are at best
potential detections. Construction of images using only long baselines
to eliminate the contribution of diffuse emission did not support
detections of MSP 18 or 188, mainly because of the significantly
smaller number of visibilities in the long-baseline range used and the
commensurate rise in the rms level of the resulting image.

If we assume that the measured fluxes are due to thermal emission from
the winds, we can estimate the corresponding mass loss rates. Assuming
the spectral types of MSP 18 and MSP 188 to be O5 III and O4 III (Rauw et
al. 2007) nad wind terminal velocities from Prinja et al. (1990), the
apparent fluxes give mass loss rates of a few $\times10^{-4}$
M$_\odot\, {\rm yr}^{-1}$ for the two stars, assuming a distance of
4--6~kpc. These are about two orders of magnitude higher than the expected
$\sim 10^{-6}$M$_\odot\, {\rm yr}^{-1}$ from standard models
(e.g. Vink et al. 2000).

Vargas \'Alvarez et al. (2013) found that MSP 188 is an O4V+O9V binary
system. Naz\'e et al. (2008) concluded that MSP 18 is assumed to be an
interacting binary. If the radio maxima at the positions of the stars
are due to emission from these binary systems, it is possible that the
flux is due to nonthermal emission arising from colliding-wind regions
between the stars, as reported for instance by Chapman et al. (1999). 

%
\begin{table*}[t]
\caption{Summary of distance estimates for RCW~49/Westerlund~2.}
\label{tabdistance}    
\centering         
\begin{tabular}{l r l l l}
\hline     
Target                 & d (kpc)          & Band/Method                   & Component             & Reference \\    
\hline  
RCW~49                 & 6                & Radio observations            & Radio nebula          & Westerlund 1961              \\
Wd2                    & 5                & Optical photometry            & Stars                 & Moffat \& Vogt 1975          \\
RCW~49                 & 0 or 4.9         & Radio recombination lines     & Radio nebula          & Caswell \& Haynes 1987       \\
Wd2                    & 7.9$^{1.2}_{1.0}$    & Optical photometry            & Stars                 & Moffat et al. 1991           \\
RCW~49                 & 2.31              &  Kinematic of HII regions     & Ionized gas            & Brand \& Blitz 1993           \\
Wd2                    & 5.7 $\pm$ 0.3    & Optical spectroscopy          & Stars                 & Piatti et al. 1998           \\
Wd2                    & 2.8              & Near-infrared photometry      & Stars                 & Ascenso et al. 2007	         \\
Wd2                    & 8 $\pm$ 1.4      & Optical spectrophotometry     & Early-type stars & Rauw et al. 2007             \\
RCW~49 and Wd2         & 2 -- 5           & X-ray photometry              & T Tauri stars	  & Tsujimoto et al. 2007        \\
Wd2, HESS J1023--575    & 6 $\pm$ 1        & CO emission, HI absorption lines & Molecular gas      & Dame 2007                    \\
RCW~49                 & 6                & IR                            & Bow shocks            & Povich et al. 2008           \\
RCW~49	               & 5.4$^{1.1}_{1.4}$ & CO emission	                  & Molecular gas         & Furukawa et al. 2009         \\
Pulsar 1023-5746       & 2.4              & $\gamma$-ray pulsar pseudo-distance & Pulsar          & Saz Parkinson et al. 2010    \\
Wd2                    & 6.5 -- 9 & Optical spectrophotometry   & Eclipsing early-type &\\ 
&&&stars + WR20 a & Rauw et al. 2011 \\
Wd2                    & 3.0          & Optical spectrophotometry / &&\\
&&anomalous extinction law & Bright stars & Carraro et al. 2013 \\
Wd2                    & 4.42$\pm$0.07    & HST spectro-photometry        & 26 O-type stars       & Vargas \'Alvarez et al. 2013\\
\hline                 
\end{tabular}
\end{table*}

\subsection{Spectral indices: Thermal vs. nonthermal emission}

As indicated in Sec.~\ref{sec:spix}, the bulk of the emission in the
region RCW~49 has a value $\sim 0$, consistent with the optically the thin
thermal plasma expected from a large HII region such as
RCW~49. However, the spectral index maps in Figs.~\ref{fig:spixwu}
and~\ref{fig:spix5-9} hint that some of the plasma in the bridge
region may have spectral index values that are significantly steeper
than -0.1, suggesting the presence of nonthermal emission due to
relativistic electrons.

We remain cautious about this potential indication due to concerns
arising from the different spatial frequency coverage at the two
frequencies used to generate the spectral index maps.  Though we have
kept the minimum baselines fixed when deriving a spectral index map,
not all baselines are covered by the observations at each frequency,
from the shortest baselines to the longest ones. Certainly, the
broad-band CABB data provide more complete baseline coverage at each
frequency and a higher degree of spatial frequency overlap between
the observing frequencies. However, a disparity in coverage at the two
frequencies remains.

Ideally, stronger evidence for the presence of nonthermal emission is
required, such as attempting to detect polarized emission in the areas
of the bridge. This will be challenging because of the high density of the
thermal plasma in this region which will Faraday-scatter any polarized
emission, possibly to a level where it is no longer detectable. 

\subsection{Nature of HESS J1023--575}

If nonthermal emission is present in the bridge, 
a simple explanation of the source HESS J1023--575 is possible.  The
bridge is very bright at IR wavelengths. Spitzer-GLIMPSE images at
the four bands (b1, b2, b3, and b4) are direct evidence that there is a
significant amount of heated dust (Churchwell et al. 2004, Whitney et
al. 2004). This medium is a target for relativistic
particles to produce high-energy radiation. The relativistic particles
in the region could be produced by the collective action of the winds
of the Westerlund~2 massive stars. The net effect is to increase the
local average cosmic ray density. 
The bright IR emission of the bridge proves to be
high-density ambient matter, which in turn will favor the interaction 
between nonrelativistic and relativistic particles, for instance, protons.
Evantually, pions are produced
from proton-proton interaction.  The neutral pions decay, giving rise
to gamma rays over an extended area (because the bridge is extended).
This extended source could in principle be identified as HESS J1023$-$575,
while 
secondary electrons would be responsible of
the synchrotron radiation. Deeper observations to increase the 
signal-to-noise ratio at better matching beams are a fundamental tool 
for confirming or discarding 
the detection of nonthermal emission from synchrotron origin 
and for testing this hypothesis.


\section{Conclusions and future work}

The broad-band ATCA CABB observations reveal a detailed view of 
radio-emitting plasma in RCW~49.  For the first time pillars of radio emission
were detected in the HII region RCW~49.
The radio emission is very similar
in morphology to the Spitzer-GLIMPSE emission, indicating that the
plasma generates free-free emission at both radio and IR wavelengths.

The brightest continuum radio knots in the bridge region have
recombination line emission. We detected the H137$\beta$ line at a
3$\sigma$ level. The H137$\beta$ line central velocity agrees very
well with that of CO (Furukawa et al. 2009). We suggest that the molecular gas
and the ionized matter are co-located at a kinematic distance of 6$\pm$1 kpc.

Spectral indices derived from the simultaneous two-frequency data showed
some hints of nonthermal radiation. We proposed an explanation for the
high-energy emission based on a proton-proton interaction scenario.
We did not detect a pulsar wind nebula around 2FGL J1022.7$-$5741 above
a threshold of 2 mJy.

A detailed study of the correlation between cm-radio data and 
high-resolution IR data (e.g. Herschel images) will allow a more detailed
description of the morphology of the plasma and shed light on the
mechanisms that excite the matter, especially the dust.  Observations
of RRLs, such as H109$\alpha$, can provide information on the
parameters of the ionized emitting gas, and also on its velocity field
and consequently its kinematic distance. Additional continuum data at
5.5 and 9 GHz will provide more sensitivity, which is required to attempt to
detect members of the Westerlund~2 cluster.

\begin{acknowledgements}
We thank the anonymous referee for the careful reading of our manuscript 
and useful comments.
PB wishes to thank people at ATNF that helped in various ways, for instance,
Nathan Pope, Marc Wieringa, Robin Wark, Vanessa Moss, Jill Rathborne,
Jim Caswell, Shari Breen, and also G.E. Romero. We are grateful 
to Yael Naz\'e,
who provided the Chandra data fits images, and to Roberta 
Paladini, who provided the results of RRL data of RCW 49 
in advance of publication. 
JM and JRSS acknowledge
support for different aspects of this work by grants
AYA2010-21782-C03-03 from the Spanish Government and Consejer\'{\i}a
de Econom\'{\i}a, Innovaci\'on y Ciencia of Junta de Andaluc\'{\i}a as
research group FQM-322, as well as FEDER funds. PB acknowledges
support from PICT 2007, 00848 (ANPCyT). This research has made use of
NASA's Astrophysics Data System Bibliographic Services, of the SIMBAD
database, operated at CDS, Strasbourg, France and of the NASA/IPAC
Infrared Science Archive, which is operated by the Jet Propulsion
Laboratory, California Institute of Technology, under contract with
the National Aeronautics and Space Administration.
\end{acknowledgements}


\section*{References} 


\noindent Abramowski, A., Acero, F., Aharonian, F., et al. 2011, A\&A, 525, 46


\noindent Aharonian, F., Akhperjanian, A. G., Bazer-Bachi, A. R., et al. 2007, A\&A, 467, 1065

\noindent Anglada, G., Villuendas, E., Estallela, R., et al. 1998, AJ, 116, 2953

\noindent Ascenso, J., Alves, J., Beletsky, Y., Lago, M. T. V. T. 2007, A\&A, 466, 137

\noindent Belloni, T. \& Mereghetti, S. 1994, A\&A, 286, 935

\noindent Benjamin, R. A., Churchwell, E., Babler, B. L., et al. 2003, PASP, 115, 953


\noindent Brand, J. \& Blitz, L. 1993, A\&A, 275, 67



\noindent Carraro, G., Turner, D., Majaess, D., Baume, G. 2013, Proceedings of the IAU Symp 289, in press (astroph:arXiv1209.2080)



\noindent Caswell, J. L. \& Haynes, R. F. 1987, A\&A, 171, 261

\noindent Chapman, J.M, Leitherer, C., Koribalski, B., et al. 1999, ApJ, 518, 890

\noindent Churchwell, E., Mezger, P. G., Huchtmeier, W. 1974, A\&A, 32, 283

\noindent Churchwell, E., Whitney, B. A., Babler, B. L. et al 2004, ApJS, 154, 322

\noindent Dame, T. M. 2007, ApJ, 665, L163

\noindent Dormody, M. for the Fermi-LAT Collaboration 2009, Fermi Symposium Conference Proceedings C091122 (astro-ph:arXiv0912.3949)

\noindent Evans, I. N., Primini, F. A., Glotfelty, K. J., et al. 2010, ApJS, 189, 37

\noindent Fujita, Y., Hayashida, K., Takahashi, H., Takahara, F. 2009, PASJ, 61, 1229

\noindent Furukawa, N., Dawson, J. R., Ohama, A., et al 2009, ApJ, 696, 115

\noindent Goss, W. M., Shaver, P. A. 1970, AuJPA, 14, 1

\noindent Johnson, K. E. 2005, IAUS 227, 413


\noindent Moffat, A. F. J. \& Vogt, N. 1975, A\&AS, 20, 155

\noindent Moffat, A. F. J., Shara, M. M., Potter, M. 1991, AJ, 102, 642


\noindent Naz\'e, Y., Rauw, G., Manfroid, J. 2008, A\&A, 483, 543

\noindent Nolan, M. C., Magri, C., Benner, L. A. M., et al. 2012, ApJS, 199, 46

\noindent Ohama, A., Dawson, J. R., Furukawa, N., et al. 2010, ApJ, 709, 975


\noindent Paladini, R., et al. 2013 (submitted)

\noindent Piatti, A. E., Bica, E., Clari\'a, J. J. 1998, A\&AS, 127, 423

\noindent Povich, M. S., Benjamin, R. A., Whitney, B. A., et al. 2008, ApJ, 689, 242

\noindent Purcell, C. R., Hoare, M. G., and the CORNISH team 2010, 
Highlights of Astronomy, 15, Ian F. Corbett, ed.



\noindent Rauw, G., Manfroid, J., Gosset, E. et al. 2007, A\&A, 436, 981

\noindent Rauw, G., Sana, H., Naz\'e, Y. 2011, A\&A, 535, 40

\noindent Reich, W. T., Rho, J., Tappe, A., et al. 2006, AJ, 131, 1479

\noindent Robinson, B. J., Brooks, J. W., Godfrey, P. D., Brown, R. D. 1974, AuJPh, 27, 856

\noindent Rodgers, A. W., Campbell, C. T., Whiteoak, J. B. 1960, MNRAS, 121, 103



\noindent Sault, R. J., Teuben, P. J., Wright, M. C. H. 1995, ASPC, 77, 433

\noindent Saz Parkinson, P. M., Dormody, M., Ziegler, M., et al. 2010, ApJ, 725, 571

\noindent Tsujimoto, M., Feigelson, E. D., Townsley, L. K., et al. 2007, ApJ, 665, 719

\noindent Vila, G. S., Aharonian, F. 2009, in ``Compact Objects and their Emission'', eds. G. E. Romero \& P. Benaglia, AAABS Vol 1, 1

\noindent Vink, J. S., de Koter, A., Lamers, H. J. G. L. M. 2000, A\&A, 362, 295

\noindent Voges, W., Aschenbach, B., Boller, Th., et al. ROSAT Faint source Cat. 2000, IAU Circ., 7432, 1

\noindent Westerlund, B. 1961, ArA, 2, 419

\noindent Whiteoak, J. B. Z. \& Uchida, K. I. 1997, A\&A, 317, 536

\noindent Whitney, B. A., Indebetouw, R., Babler, B., L. et al. 2004, ApJS, 154, 315

\noindent Wilson, W. E., Ferris, R. H., Axtens, P. et al. 2011, MNRAS, 416, 832




\end{document}